\documentclass[aps,prl,twocolumn,showpacs,preprintnumbers,amssymb]{revtex4}

\usepackage{graphicx}

\let\bm=\bibitem

\def\fft#1#2{{#1 \over #2}}

\def\nn{\nonumber}
\def\sst#1{{\scriptscriptstyle #1}}
\def\0{{\sst{(0)}}}
\def\1{{\sst{(1)}}}
\def\2{{\sst{(2)}}}
\def\3{{\sst{(3)}}}
\def\4{{\sst{(4)}}}
\def\5{{\sst{(5)}}}
\def\6{{\sst{(6)}}}
\def\7{{\sst{(7)}}}
\def\8{{\sst{(8)}}}
\def\ep{{{\epsilon}}}

\newcommand{\bea}{\begin{eqnarray}}
\newcommand{\eea}{\end{eqnarray}}
\newcommand{\be}{\begin{equation}}
\newcommand{\ee}{\end{equation}}

\def\Bbb{\mathbb}

\begin{document}

\preprint{DAMTP-2005-60\ \ \ \ MIFP-05-15\ \ \ \ {\bf hep-th/0507034}}

\title{AdS/CFT Casimir Energy for Rotating Black Holes}

\author{G.W. Gibbons$^1$, M.J. Perry$^1$ and C.N. Pope$^{2}$}
\altaffiliation[]{Research supported in part
by DOE grant DE-FG03-95ER40917 and NSF grant INTO3-24081.}
\affiliation{%
${}^1\!\!\!$ DAMTP, Centre for Mathematical Sciences, Cambridge University
Wilberforce Road, Cambridge CB3 OWA, UK\\
${}^2\!\!\!$ George P. \& Cynthia W.
Mitchell Institute for Fundamental Physics,
Texas A\&M University, College Station, TX 77843-4242, USA
}%

\date{July 4, 2005}

\begin{abstract}

   We show that if one chooses the Einstein Static Universe as the
metric on the conformal boundary of Kerr-AdS spacetime, then the
Casimir energy of the boundary conformal field theory can easily be
determined.  The result is independent of the rotation parameters, and
the total boundary energy then straightforwardly obeys the first law
of thermodynamics. Other choices for the metric on the conformal
boundary will give different, more complicated, results. As an
application, we calculate the Casimir energy for free self-dual tensor
multiplets in six dimensions, and compare it with that of the
seven-dimensional supergravity dual. They differ by a factor of $5/4$.

\end{abstract}

\pacs{11.25.-w, 98.80.Jk, 04.50.+h}
\maketitle

    An interesting application of the AdS/CFT correspondence 
\cite{mald,guklpo,wit} is to consider for the bulk solution
a rotating Kerr-anti-de Sitter spacetime.  As was discussed in
\cite{hawhuntay}, the boundary theory in this case should describe a
conformal field theory in a rotating Einstein universe, allowing one in
principle to study the effects of rapid rotation on the thermodynamics of
the system.  Of particular interest is the behaviour of the conformal
field theory (CFT) on the boundary as the rotation parameters achieve their
maximal values consistent with the non-existence of closed timelike 
curves.  The Chronology Protection Conjecture of Hawking \cite{hawk} suggests
that going beyond this limit should be physically impossible.  This may
be associated with a divergence of the free energy of the CFT as one
approaches the limit, and with a possible non-unitarity of the CFT if one
passes beyond it.

   The study of the thermodynamics of such rotating systems is quite
involved and subtle, both in the bulk, and in the
passage to the boundary theory.  In the bulk, there
are subtleties concerning the definition of the energy, or mass, of the
rotating AdS black hole.  As we showed in \cite{gibperpop}, it is important
that one evaluate the energy and the angular velocities with respect to a 
frame that is 
asymptotically non-rotating at infinity, in order to obtain quantities 
that satisfy the first law of thermodynamics,
\be
dE = T dS + \Omega^i\, dJ_i\,.\label{firstlaw}
\ee
In particular, a commonly considered frame that rotates relative to
the asymptotically static frame, and which arose when the Kerr-AdS
metrics were first constructed in Boyer-Lindquist type coordinates, 
is particularly inappropriate for defining the energy and angular 
velocities, since its asymptotic rotation rate is dependent on the 
angular-momentum parameters of the metric \cite{gibperpop2}.

   In \cite{gibperpop2} we addressed the question of how one should
map between the bulk thermodynamic variables and the corresponding 
variables on the boundary.  We showed that using the standard UV/IR
connection, the bulk variables $(E,T,S,\Omega^i,J_i)$ 
that satisfy the first law as in 
(\ref{firstlaw}) map straightforwardly into boundary variables 
$(E,T,S,\omega^i, j_i)$ that
also satisfy the first law, now with the addition of a pressure term,
\be
de= t ds + \omega^i dj_i - p dv\,.
\ee
This mapping is implemented, for an $n$-dimensional bulk spacetime,
by imposing the relations 
\bea
&&e= \fft{l}{y}\, E\,,\qquad \omega^i = \fft{l}{y}\, \Omega^i\,,\qquad
t = \fft{l}{y}\, T\,,\qquad
s=S\,,\nn\\
&& j_i= J_i\,,\qquad 
v = {\cal A}_{n-2}\,  y^{n-2}\,,\qquad p = \fft{e}{(n-2)v}\,.
\eea
Here $l$ is the radius of the asymptotically-AdS spacetime, in the
sense that $R_{\mu\nu} = - (n-1) l^2 g_{\mu\nu}$, ${\cal A}_{n-2}$ is
the volume of the unit $(n-2)$-sphere, and $y$ is the radius of a
large sphere, with round spherical metric, near the boundary in AdS.
As we emphasised in \cite{gibperpop2}, the natural choice of
boundary metric is that of the Einstein Static Universe, ESU$_{n-1}$
i.e. the standard product metric on ${\Bbb R} \times S^{n-2}$. In
other words, one introduces a set of coordinates in which the Kerr-AdS
metric approaches the canonical AdS$_n$ metric at infinity, in the
form
\bea
d\bar s^2_n &=& - (1+ y^2 l^{-2}) dt^2 + \fft{dy^2}{1+ y^2 l^{-2}}
+ y^2 \, d\Omega_{n-2}^2\,,
\label{boundmet}
\eea
where $d\Omega_{n-2}^2$ is the metric on the
unit $(n-2)$-sphere.   
The conformal boundary is then located at $y\rightarrow \infty$,
and the induced metric has the standard ESU$_{n-1}$ form
\be
d\bar s^2_{n-1}= -dt ^2 + 
   l^2\, d\Omega_{n-2}^2 \,.\label{esumet}
\ee
More precisely, in an AdS type conformal compactification,
the physical metric is given by $g = \Omega ^{-2} \tilde g$,
where $\Omega =0$ and $d \Omega \ne 0$
on the timelike conformal boundary $\cal I$, with $\tilde g$ being 
non-singular in a neighbourhood of $\cal I$. The metric induced
on the boundary $\cal I$ is $h=\tilde g|_{\cal I}$,
and depends on the choice of the conformal factor $\Omega$.
If $\Omega \rightarrow f \Omega$, for some 
function $f$ which is non-vanishing in a neighbourhood of $\cal I$,
then the boundary metric undergoes a conformal rescaling
 $h \rightarrow f^2 h$. In order to obtain the standard 
metric on ESU$_{n-1}$, one simply chooses $\Omega ={l \over y}$.
  
     Our result in \cite{gibperpop2} refuted a recent surprising claim
in \cite{cai}, where it was asserted that in order to get
thermodynamic quantities that satisfied the first law on the boundary,
it was necessary to start from thermodynamic quantities in the bulk
that did not satisfy the first law (in fact, the quantities defined
relative to the rotating frame we mentioned earlier).  As we showed,
the puzzling results in \cite{cai} were associated with the use of a
somewhat unnatural conformal factor $\Omega$, and hence a complicated
boundary metric $h$ whose spatial sections were not round spheres, and
which was not ultrastatic, i.e.  $g_{tt}$ depended on spatial
position.

   In \cite{gibperpop2}, our principal interest was in the properties and
thermodynamics of the bulk theories, and so it was  not relevant to   
consider the contribution of Casimir energies on the boundary.  Such 
terms do play an important r\^ole in the boundary CFT, and much work
has been done on calculating them.  For the case of Schwarzschild-AdS 
spacetime,
the boundary Casimir contribution was evaluated in \cite{balakrau}.  
Casimir calculations have also been performed for the case of rotating  
Kerr-AdS black holes, in \cite{awadjohn} and more recently in
\cite{skenpapa}.  The result obtained in these papers for the 
four-dimensional boundary theory is
\be
E_{\rm Casimir} = \fft{3\pi^2 l^2}{4\kappa^2}\, 
  \Big(1 + \fft{(\Xi_a-\Xi_b)^2}{9 \Xi_a \Xi_b}\Big)\,,\label{ecasimir}
\ee
where $\kappa^2/(8\pi)$ is Newton's constant, 
$\Xi_a= 1 - a^2 l^{-2}$, $\Xi_b=1 - b^2 l^{-2}$, and $a$ and $b$
are the rotation parameters of the five-dimensional Kerr-AdS black hole
given in \cite{hawhuntay}.  

   The expression (\ref{ecasimir}) for the Casimir energy of the
boundary is a somewhat surprising result, since it depends on the
rotation parameters $a$ and $b$ of the Kerr-AdS black hole.  As we
have argued above, the most natural conformal frame in which to
formulate the boundary CFT is one in which the metric approaches the
form (\ref{boundmet}), and the boundary metric is that of ESU$_{n-1}$,
and since this metric is manifestly independent of $a$ and $b$ the
Casimir energy will necessarily also be independent of $a$ and $b$.
Evidently, therefore, the conformal boundary metric chosen in
\cite{awadjohn,skenpapa} is not the one we are advocating.  In the
remainder of this paper, we shall endeavour to convince the reader
that our proposed choice of conformal boundary metric for the CFT dual
to the Kerr-AdS metric is by far the most natural one, and that it has
the very satisfactory feature that it leads to a genuinely constant
Casimir energy in the boundary theory.

   In fact, in order to demonstrate our point we need only collect a few
results from previous papers.  We shall mostly discuss the general 
$n$-dimensional case, since it is just as easy to discuss it generally 
as in any specific dimension.  The general Kerr-AdS metrics were obtained in
\cite{gilupapo1,gilupapo2}. 
The metrics have $N\equiv [(n-1)/2]$ independent rotation
parameters $a_i$ in $N$ orthogonal 2-planes.  We have $n=2N+1$ when
$n$ is odd, and $n=2N+2$ when $n$ is even.  The metrics can be described by
introducing $N$ azimuthal angles $\phi_i$, and $(N+\ep)$ ``direction
cosines'' $\mu_i$ obeying the constraint
\be
\sum_{i=1}^{N+\ep} \mu_i^2 =1\,,\label{muconstraint}
\ee
where $\ep= (n-1)$ mod 2.
In Boyer-Lindquist type coordinates that are asymptotically non-rotating,
the metrics are given by \cite{gilupapo1,gilupapo2}
\bea
ds^2_n  &=&  - W\, (1 + r^2\, l^{-2} )\, dt^2
+ \sum_{i=1}^N \fft{r^2 + a_i^2}{\Xi_i}\,\mu_i^2\,
    d\varphi_i^2 \nn\\
&&
 + \sum_{i=1}^{N+\ep} \fft{r^2 + a_i^2}{\Xi_i}\, d\mu_i^2
 + \fft{U\, dr^2}{V-2m}\nn\\
&&
 + \fft{2m}{U}\Bigl(W\,dt
 - \sum_{i=1}^N \fft{a_i\, \mu_i^2\, d\varphi_i}
  {\Xi_i }\Bigr)^2\nn \nn\\
&&
 - \fft{l^{-2}}{W\, (1 + r^2\, l^{-2})}
    \Bigl( \sum_{i=1}^{N+\ep} \fft{r^2 + a_i^2}{\Xi_i}
    \, \mu_i\, d\mu_i\Bigr)^2 \,,\label{bl}
\eea
where
\bea
W &\equiv& \sum_{i=1}^{N+\ep} \fft{\mu_i^2}{\Xi_i}\,,\qquad
 \Xi_i\equiv 1 - a_i^2\, l^{-2}\,,\nn\\
U &\equiv & r^{\ep}\, \sum_{i=1}^{N+\ep} \fft{\mu_i^2}{r^2 + a_i^2}\,
\prod_{j=1}^N (r^2 + a_j^2)\,,\\
V &\equiv& r^{\ep-2}\, (1 +r^2\, l^{-2})\,
   \prod_{i=1}^N (r^2 + a_i^2)\,,
\label{uvw}
\eea
and it is understood, in the even-dimensional case, that $a_{N+1}=0$.
The metrics satisfy $R_{\mu\nu}=-(n-1)\, l^{-2}\, g_{\mu\nu}$.
The constant-$r$ spatial surfaces at large distance are inhomogeneously
distorted $(n-2)$-spheres.  Making the coordinate transformations
\be
\Xi_i \,  y^2 \, \hat \mu_i^2 =(r^2 +a_i^2)\, \mu_i^2\,,\label{ctrans}
\ee
where $\sum_i \hat\mu_i^2=1$, the metrics at large $y$ approach the
standard AdS form given in (\ref{boundmet}), where
\be 
d\Omega_{n-2}^2 = \sum_{k=1}^{N+\ep} d\hat\mu_k^2 + \sum_{k=1}^N
\hat\mu_k^2  d\varphi_k^2\,,\label{sphmet}
\ee
with round $(n-2)$-spheres of volume ${\cal A}_{n-2}\, y^{n-2}$ at
radius $y$, where ${\cal A}_{n-2}$ is the volume of the unit $(n-2)$-sphere.
Note, in particular, that the boundary metric does not depend on any of
the black-hole parameters.

   The boundary CFT will be defined on the surface $y=\,$constant at
very large $y$. The Casimir energy in the boundary theory is clearly
independent of the mass $m$ of the Kerr-AdS black
hole, since the boundary metric does not depend upon $m$.  
Therefore, for convenience, we can evaluate it by first setting $m=0$ in
(\ref{bl}), which implies that the metric becomes purely AdS itself, in an
unusual coordinate system.  As was shown in
\cite{gilupapo1,gilupapo2}, if one now performs the coordinate
transformation (\ref{ctrans}), the AdS metric becomes {\it exactly}
the canonical metric given in (\ref{boundmet}).  The calculation of
the Casimir energy for the boundary of Kerr-AdS is therefore reduced
to the standard calculation for the Einstein Static Universe,
ESU$_{n-1}$.  The answer is obviously independent of the rotation
parameters, since they do not appear in the boundary metric.

   One may calculate the Casimir energy in a number of ways.  In
\cite{balakrau}, it was shown that the use of the holographic stress
tensor for the conformal anomaly for four-dimensional ${\cal N}=4$
super-Yang-Mills gives
\be
E_{\rm Casimir} = \fft{3\pi^2 l^2}{4\kappa^2}\,.\label{ecasimir2}
\ee
This value agrees with a direct calculation of the Casimir energy
using zeta-function regularisation of the sums over energy eigenvalues
for $6N^2$ scalar fields, $4N^2$ Weyl spinor fields and $N^2$ vector
fields on $S^3$.  For our choice of conformal boundary metric,
(\ref{ecasimir2}) is therefore the Casimir energy in the
four-dimensional boundary theory dual to the five-dimensional Kerr-AdS
metric.  Combined with our calculation of the bulk energy term
obtained in \cite{gibperpop}, the complete CFT energy is given by
\be
E_{\rm tot} = \fft{2\pi^2 m(2\Xi_a + 
          2\Xi_b - \Xi_a \Xi_b)}{\kappa^2 \Xi_a^2 \Xi_b^2} +
  \fft{3\pi^2 l^2}{4\kappa^2}\,.
\ee

  The calculations in any other dimension proceed similarly, again yielding
Casimir energies that are necessarily independent of the black hole 
rotation parameters.  For example, in the case of a seven-dimensional 
Kerr-AdS bulk spacetime, we obtain a total boundary energy given by 
\be
E_{\rm tot} = \fft{2 m \pi^3}{\kappa^2\, (\prod_i \Xi_i)}\, 
\Big(\fft{1}{\Xi_1} + \fft1{\Xi_2} + \fft1{\Xi_3} - \fft1{2}\Big) 
   - \fft{5 \pi^3 l^4}{16 \kappa^2}\,,\label{d7en}
\ee
where the first term is the bulk energy that we calculated in
\cite{gibperpop}, and we have read off the Casimir term by setting
$a=0$ in equation (51) of \cite{awadjohn}.  This value, which came
from the use of the holographic stress tensor, should be compared with
the value obtained directly by zeta-function regularisation of the
sums over energy eigenvalues of $20N^3$ scalars, $8N^3$ Weyl fermions
and $4N^3$ self-dual 3-forms, making up $4N^3$ copies \cite{basfrotse}
of the $(2,0)$ tensor multiplet.  Using the energies and degeneracies
tabulated in \cite{kutlar}, we find that the Casimir energy for $N_0$
scalars, $N_{1/2}$ Weyl fermions and $N_T$ self-dual 3-forms is
\be
E_{\rm Casimir}= -\fft{(124 N_0  + 1835 N_{1/2} +
       11460 N_T)}{241920\, l}\,.
\ee
For $4N^3$ multiplet, with the seven-dimensional AdS/CFT relation 
$N^3=3\pi^3 l^5/(2\kappa^2)$, this gives
\be
E_{\rm Casimir} = -\fft{25 \pi^3 l^4}{64 \kappa^2}\,.
\ee
Clearly this does not agree with the Casimir term in (\ref{d7en}); the
ratio is in fact $5/4$, which is not the same as the $4/7$ ratio
conjectured in \cite{awadjohn}.   Presumably these differences reflect
the absence of the non-renormalisation theorems that appear to account
for the agreement of the various methods in the four-dimensional case.

     We have demonstrated that by choosing the conformal boundary
metric defined by taking $y=\,$constant, we obtain a very simple
framework for describing the thermodynamics of the boundary field
theory, with a very simple expression for the Casimir contribution to
the energy, which is independent of the parameters in the Kerr-AdS
metric.  In particular, this means that the first law of
thermodynamics continues to hold in a straightforward manner, when the
Casimir energy is included.  This contrasts with the more complicated
situation in the case of the conformal boundary metric chosen in
\cite{awadjohn,skenpapa}, where, as was shown in \cite{skenpapa}, an
additional diffeomorphism term must be included in the first law in
order to compensate for the dependence of the boundary metric on the
rotation parameters of the black hole.

   The Boyer-Lindquist form of the boundary metric (in non-rotating
coordinates) is obtained by taking the limit $r\rightarrow\infty$ of
$(l^2/r^2)\, ds^2$, where $ds^2$ is given by (\ref{bl}).  This
yields
\bea
ds^2_{n-1} &=& -W dt^2 + l^2 \sum_{i=1}^{N+\ep} \fft1{\Xi_i}\, d\mu_i^2
  + l^2 \sum_{i=1}^N \fft{1}{\Xi_i}\, \mu_i^2 d\varphi_i^2 \nn\\
&&-
   \fft{l^2}{W}\, \Big(\sum_{i=1}^{N+\ep} \fft{\mu_i d\mu_i}{\Xi_i}\Big)^2
\,.\label{blbound}
\eea
If we introduce new coordinates $\hat \mu_i$, 
satisfying $\sum_i \hat\mu_i^2=1$, by the transformations
\be
\mu_i = \sqrt{W \Xi_i}\, \hat\mu_i\,,
\ee
then the metric (\ref{blbound}) becomes
\be
ds_{n-1}^2 = W\, d\bar s_{n-1}^2\,,
\ee
where $d\bar s_{n-1}^2$ is the standard metric on ESU$_{n-1}$, given
by (\ref{esumet}) and (\ref{sphmet}), with $W$ expressed in terms of
the $\hat\mu_i$ as $W=\sum_i \Xi_i\, \hat\mu_i^2$.  This is the
generalisation to arbitrary dimension of the five-dimensional
demonstration in \cite{gibperpop} that the $r=\,$constant
``Boyer-Lindquist'' boundary metric (\ref{blbound}) and the
$y=\,$constant Einstein static universe boundary metric (\ref{esumet})
are related by a Weyl rescaling, and thus they lie in the same
conformal class.  When considering the first law, one varies the
rotation parameters $a_i$.  If one uses the Boyer-Lindquist boundary
metric (\ref{blbound}), this variation induces an infinitesimal change
of the conformal factor, and hence a change of the Casimir
contribution to the energy, as described in detail in five dimensions
in \cite{skenpapa} (see their equations (6.51)-(6.54)).

   We are not, of course, saying that the more complicated results
for the Casimir energies obtained in \cite{awadjohn,skenpapa} are
wrong, but rather, that they are the consequence of having made a less
felicitous choice for the conformal boundary metric, or equivalently,
the conformal factor.  As one can see from the calculations presented
in \cite{skenpapa}, although a coordinate transformation was performed
in order to simplify the conformal boundary metric, it did not lead to
as great a simplification as can be achieved by using the transformation
(\ref{ctrans}).  

    We shall conclude with a remark on the action of the asymptotic
symmetry group.  Since we are working on the universal covering space
$\widetilde{\hbox{AdS}}_n$ of AdS$_n \equiv SO(n-1,2)/SO(n-2,1)$, this is
an infinite covering ${\widetilde {SO}} (n-1,2)$ of $SO(n-1,2)$.  The bulk
energy $E$ and the angular momenta $J_i$ transform properly under the
action of the asymptotic symmetry group as the components of a
$5\times 5$ antisymmetric tensor $J_{AB}$, which may be thought of as
an element of the Lie algebra ${\widetilde{\mathfrak{so}}}(n-1,2)$. The
transformation rule is just the adjoint action.  

    The transformation properties of the Casimir energy are more
subtle. In general, the asymptotic group acts on the boundary by
consometries, i.e. preserving the boundary metric $h$
only up to a conformal factor. Depending upon one's choice of the
boundary metric, i.e, of the representative in its conformal
equivalence class, there may be a subgroup which acts by
isometries. If, as we have done, we choose the Einstein static
universe ESU$_{n-1}$ as boundary metric, then this subgroup is ${\Bbb
R}\times SO(n-1)$ and is maximal.  Indeed, it is an infinite covering
of the maximal compact subgroup of $SO(n-1,2)$. The Casimir
contribution to the energy is clearly invariant under the subgroup of
isometries of the boundary metric, but it transforms in a well-defined
but more complicated and non-trivial fashion under those elements of
the asymptotic symmetry group which do not induce isometries.
Choosing the Einstein static universe ESU$_{n-1}$ as boundary metric
minimises these complications.

   In summary, we have seen from previous work that the description of
the bulk thermodynamics of rotating black holes in an AdS background is
greatly simplified if one refers the energy and the angular velocities
of the black hole to a coordinate frame that is non-rotating at infinity.
Especially, one should not choose a rotating coordinate frame whose 
asymptotic angular velocity depends upon the parameters of the black hole.
In this paper, we have furthermore shown that the description of the
boundary CFT is greatly simplified if one likewise defines a conformal boundary
metric that does not depend upon the parameters of the black hole.
This is straightforwardly achieved by applying the coordinate
transformation (\ref{ctrans}) to the asymptotically-static form of
the Kerr-AdS metrics given in (\ref{bl}), and defining the boundary metric
as the section $y=\,$constant as $y$ tends to infinity.  By this means, all the
complications associated with the unnecessary introduction of black 
hole parameter dependence of the thermodynamical quantities in the 
boundary theory are avoided.


\begin{acknowledgments}


   We thank Adel Awad, David Berman, Clifford Johnson and Ioannis
Papadimitriou for discussions.  C.N.P. thanks the Relativity and
Cosmology group, Cambridge, for hospitality during the course of this
work.

\end{acknowledgments}

\end{document}